\documentclass[10pt]{iopart}

\usepackage{iopams}  
\usepackage{graphicx}
\usepackage{amssymb}
\usepackage{epstopdf}
\usepackage{float}
\usepackage{color}
\usepackage{subfigure}
\usepackage{multirow}
\usepackage{relsize}
\usepackage{booktabs}
\expandafter\let\csname equation*\endcsname\relax

\expandafter\let\csname endequation*\endcsname\relax

\usepackage{amsmath}
\usepackage[ngerman, english]{babel}

\begin{document}

\title{Mapping trilobite state signatures in atomic hydrogen }

\author{Jes\'{u}s P\'{e}rez-R\'{i}os$^1$, Matthew T Eiles$^1$, Chris H Greene$^{1,2}$}

\address{$^1$Department of Physics and Astronomy, Purdue University, West Lafayette, Indiana, 47907, USA.\newline
$^2$ Purdue Quantum Center, Purdue University, West Lafayette, Indiana, 47907, USA.}
\ead{\mailto{jperezri@purdue.edu*},\mailto{meiles@purdue.edu},\mailto{chgreene@purdue.edu}}

\vspace{10pt}
\begin{indented}
\item \today
\\
*corresponding author
\end{indented}

\begin{abstract}
A few-body approach relying on static line broadening theory is developed to treat the spectroscopy of a single Rydberg excitation to a trilobite-like state immersed in 
a high density ultracold medium. The present theoretical framework implements the recently developed compact 
treatment of polyatomic Rydberg molecules, allowing for an accurate 
treatment of a large number of perturbers within the Rydberg orbit. This system exhibits two unique spectral signatures: its lineshape depends on the Rydberg quantum number $n$ but, strikingly, is independent of the density of the medium, and it is characterized by sharply peaked features reflecting the oscillatory structure of the potential energy landscape. 

\end{abstract}

%

%
%
%
\ioptwocol

\newcommand{\be}{\begin{equation}}
\newcommand{\ee}{\end{equation}}
\newcommand{\lap}{\nabla^2}
\newcommand{\pd}[1]{\frac{\partial}{\partial{#1}}}
\newcommand{\pdd}[1]{\frac{\partial^2}{\partial{#1}^2}}
\newcommand{\pdde}[2]{\frac{\partial^2{#1}}{\partial{#2}^2}}
\newcommand{\pde}[2]{\frac{\partial{#1}}{\partial{#2}}}
\newcommand{\ar}{(\vec{r},t)}
\newcommand{\arz}{(\vec{r},0)}
\newcommand{\psirt}{\psi(\vec{r},t)}
\newcommand{\psirz}{\psi(\vec{r},0)}
\newcommand{\psirzc}{\psi^*(\vec{r},0)}
\newcommand{\psirtc}{\psi^*(\vec{r},t)}
\newcommand{\probcur}{\vec{S}\ar}
\newcommand{\twodvec}[2]{\left[\begin{array}{cc}{#1} \\ {#2}\end{array}\right]}
\newcommand{\twodmat}[4]{\left[\begin{array}{cc}{#1} & {#2} \\ {#3} & {#4}\end{array}\right]}
\newcommand{\threedvec}[3]{\left[\begin{array}{ccc}{#1} \\ {#2} \\ {#3}\end{array}\right]}
\newcommand{\threedmat}[9]{\left[\begin{array}{ccc}{#1} & {#2}& {#3} \\ {#4} & {#5} & {#6}\\ {#7} & {#8} & {#9}\\\end{array}\right]}
\newcommand{\fourdvec}[4]{\left[\begin{array}{cccc}{#1} \\ {#2} \\ {#3} \\ {#4}\\ \end{array}\right]}
\newcommand{\fourdmat}[4]{\left[\begin{array}{cccc}{#1}\\{#2}\\{#3}\\{#4}\\\end{array}\right]}
\newcommand{\bra}[1]{\langle{#1}|}
\newcommand{\ket}[1]{|{#1}\rangle}
\newcommand{\bkt}[2]{\langle{#1}|{#2}\rangle}
\newcommand{\apdag}{a_+^\dagger}
\newcommand{\amdag}{a_-^\dagger}
\newcommand{\am}{a_-}
\newcommand{\ap}{a_+}
\newcommand{\sol}{\textbf{Solution: }}
\newcommand{\sumn}{ \sum_{n = 0}^N}
\newcommand{\com}[2]{\left[{#1},{#2}\right]}
\newcommand{\h}{\frac{1}{2}}
\newcommand{\hh}{\frac{1}{2}}
\newcommand{\dd}[1]{\mathrm{d}{#1}}
\newcommand{\ddd}[1]{\mathrm{d}^3{#1}}
\newcommand{\ddn}[2]{\mathrm{d^{#1}}{#2}}
\newcommand{\up}{\ket{\uparrow}}
\newcommand{\dn}{\ket{\downarrow}}
\newcommand{\upb}{\bra{\uparrow}}
\newcommand{\dnb}{\bra{\downarrow}}
\newcommand{\vv}[1]{\underline{#1}}
\newcommand{\bigO}{\mathcal{O}}
\newcommand{\del}{\underline \nabla}
\newcommand{\+}[1]{\ensuremath{\mathbf{#1}}}
\newcommand*\rfrac[2]{{}^{#1}\!/_{#2}}

 The spectroscopic study of Rydberg atoms in a background gas originated 
 in the outstanding work of Amaldi and Segr\`{e} in 1934, who reported a change in the line shape and broadening due 
 to the density and nature of the atomic background gas \cite{Segre}. Shortly afterwards, Fermi explained these intriguing observations in terms of collisions between the 
 Rydberg electron and neutral atoms (``perturbers'') within its orbit \cite{Fermi}. This theoretical framework 
 predicts the broadening and line shape at a given density to be independent 
 of the Rydberg quantum number $n$. Recently, it has become possible to study the 
 spectroscopy of a single Rydberg atom in a high-density environment 
 \cite{PfauBEC,GajNatComm}, which has revealed unexpected features of the line 
 shape beyond Fermi's original framework, such as an $n$-dependent shift of the spectroscopic 
 lines \cite{Schlag}.

The interaction of the Rydberg electron with a perturber significantly alters the 
energy landscape of the Rydberg atom. When the electron-perturber interaction 
is characterized by a negative scattering length this interaction can lead to the formation of long-range
molecular states with permanent dipole moments, the so-called trilobite states
 \cite{GreeneSadeghpourDickinson}. These exotic molecules are formed by the mixture of 
 nearly degenerate high angular momentum states, which leads to an oscillating potential 
 energy curve (PEC) extending to thousands of Bohr radii \cite{Granger}. These molecular states have 
 recently been observed in a thermal gas of Cs \cite{Shaffer}. Another exotic state occurs 
 when the electron-perturber interaction possesses a $p$-wave scattering resonance, which then produces a different type of molecular state known as a 
 butterfly Rydberg molecule \cite{HamiltonGreeneSadeghpour}. Very recently this state has been observed in Rb \cite{PendularButterflies}.

 Two different approaches have been developed to explain the unique features of the 
 Rydberg spectral lines when a single $nS$ rubidium Rydberg atom is immersed in a dense background gas. One approach 
 describes the line shape in terms of the many-body effects of the background 
 gas \cite{Schmidt2015}, whereas the second explains the same physical observables from a 
 few-body perspective by employing the static approximation of line broadening 
 theory \cite{Schlag,Stellar}. The many-body approach explicitly includes correlation 
 effects within the background gas atoms at finite temperature, leading 
 to a fairly satisfactory explanation of some aspects of the observed line shapes \cite{Schmidt2015}. On the other hand,
  the quasi-static few-body approach describes other details of the observed line shapes fairly well by sampling the energy shifts in the frozen nuclear Rydberg levels associated with different configurations of the perturbers \cite{Schlag}. This approach readily shows the effect of the 
 Rydberg-neutral PEC in the observed line shapes, and hence it can be a robust 
 technique to probe and test different Rydberg-neutral PECs within the range of validity of the quasi-static approximation. 
 
The present study explores the spectroscopic line shapes that should be observable for a hydrogen Rydberg atom immersed in a 
high-density hydrogen background gas. Neglecting corrections due to the spin-orbit interaction, hyperfine splittings, and the Lamb shift, the $nl$ states of hydrogen are degenerate. The present approach therefore unifies the few-body 
approach of M. Schlagm\"{u}ller et al. \cite{Schlag} with the study of polyatomic trilobite states \cite{ours}. This requires the inclusion of non-pairwise-additive three, four, etc. $N$-body interactions, which were treated only in a pairwise approximation for the $nS$ states of Rb \cite{GajNatComm,Schlag}.  Although in hydrogen the triplet and singlet $e^-$-H scattering lengths are both positive and therefore no weakly bound Rydberg molecule states form, we nonetheless predict the existence of sharp spectral features at $n$-dependent detunings even in the dissociative continuum. Dense hydrogen clouds are attainable in both Bose-Einstein condensate (BEC) \cite{HydrogenBECPRLs} and thermal cloud regimes; throughout this paper we present results obtained only for a thermal cloud since these results compare very favorably with identical calculations in the BEC regime.
 
Fermi's approach simplifies the interaction between the Rydberg electron and a neutral atom 
to a contact delta-function potential proportional to the scattering length; this is known as the 
Fermi pseudopotential \cite{Fermi}. This pseudopotential has been used to study Rydberg-neutral interactions with considerable success \cite{DuGreene,DePrunele}. Extending this model to a Rydberg hydrogen atom in 
the presence of $N$ neutral hydrogen atoms placed at positions $\vec R_i$ within the 
Rydberg blockade radius gives a potential energy (in atomic units) \cite{Rost2006}:

\be
\label{eq-1}
V = 2\pi \sum_{i=1}^{N}A_{S,T}(R_i)\delta^3(\vec r - \vec R_i),
 \ee 

\noindent
where  $A_{S,T}(R_i)$ denotes the energy-dependent singlet ($S$) and 
triplet ($T$) scattering lengths \cite{Tarana}:

\begin{align*}
A_{S,T}(R) &= A_{S,T}^0 \left[1 - \frac{\pi\alpha_d k}{3A_{S,T}^0} - \frac{4\alpha_d}{3}k^2\ln\left(\frac{\sqrt{\alpha_d}k}{4}\right)\right]^{-1}.
\end{align*}

\noindent
The semiclassical momentum is $[k(R)]^2 = 2/R-1/n^{2}$ and the static polarizability of 
hydrogen is $\alpha_d = 9/2$. The scattering lengths at  zero energy are $A_{S}^0$ = 5.965 a$_{0}$ and $A_{T}^0$ = 1.7686 a$_{0}$ \cite{Schwartz}. 
The present study concentrates on triplet spin states because they can be readily explored in a spin-polarized BEC.  Previous work has shown that this pseudopotential agrees qualitatively with more sophisticated potentials at low $n$, and its accuracy increases to a quantitative level for higher Rydberg levels \cite{DuGreene,DePrunele,Tarana,GadeaDickinson}.

\begin{figure}[t]
\begin{center}
\hspace{-1.3cm}
\includegraphics[scale=0.35]{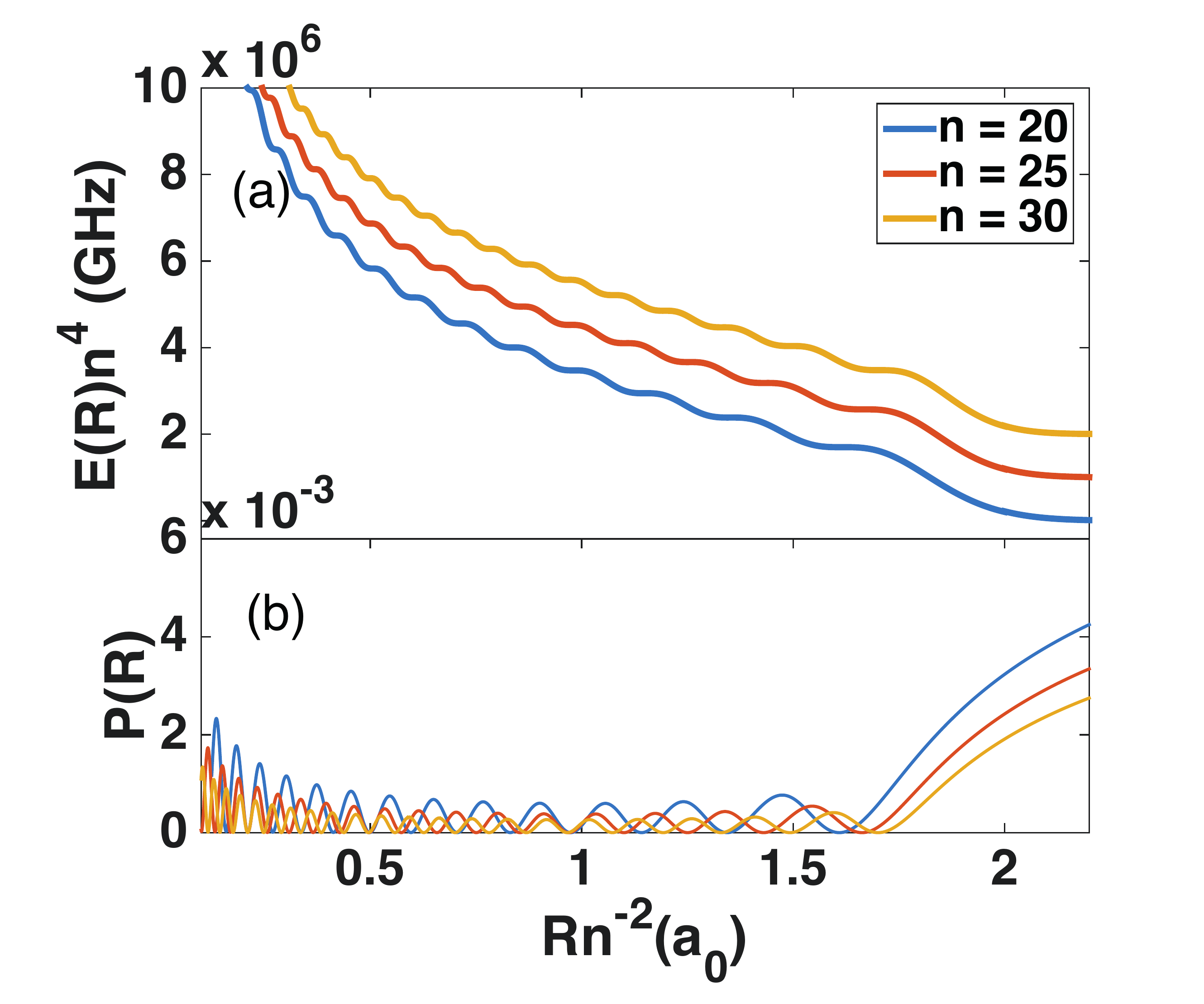}
\vspace{-20pt}
\caption{ Panel (a): diatomic potential energy curves $E(R)$ in GHz for different triplet Rydberg states 
of hydrogen, $H(n) + H(1s)$, plotted in scaled coordinates and spaced vertically by $10^6$ in these scaled units for clarity. The step-like structure of these potentials similarly arises in calculations of singlet $H(n)+H(1s)$ potential energy curves \cite{Tarana}. Panel (b): probability of two-photon excitation $\mathcal{P}(R)=|\bkt{nD}{\Psi_T}P^n_D + \bkt{nS}{\Psi_T}P^n_S|^2$, integrated over angular degrees of freedom  as a function of the 
distance (see text for details) for the same Rydberg states of panel (a). }
\label{Matt_figure}
\end{center}
\end{figure}

Treating the interaction potential as a perturbation of the degenerate hydrogenic 
states $\psi_{nlm}(\vec r) = \frac{u_{nl}(r)}{r}Y_{lm}(\hat r)$ with principal Rydberg quantum number $n$ and orbital angular momentum $l$ with 
projection $m_{l}$, the energy shift can be obtained in standard degenerate perturbation theory. This becomes calculationally cumbersome since 
the number of states scales as $n^2$. However, the 
first order normalized perturbed eigenstates - the trilobite states - of a delta function potential are analytically known:
 \begin{align}
 \label{trilobite}
 \Psi(\vec R_i,\vec r) &=\frac{1}{\mathcal N} \sum_{l}\frac{u_{nl}(R_i)}{R_i}\frac{u_{nl}(r)}{r}\frac{2l+1}{4\pi}P_l(\hat R_i\cdot\hat r);\\
 \mathcal{N}^2 &= \Psi(\vec R_i,\vec R_i),\nonumber
 \end{align}
 
 \noindent
and have been employed in a recent approach as a reduced basis to diagonalize 
Eq. (\ref{eq-1}) \cite{ours}. This reduces the dimensionality of the problem to at most $N$, the number of 
perturbers, and is exact for hydrogen within the approximation of degenerate perturbation
 theory. This leads to a generalized eigenvalue problem 
  \begin{eqnarray}
 \label{eq-2}
 {\bf A}\vec a &=E{\bf B}\vec{a}; \nonumber \\
{\bf A_{\alpha \beta}}&=2\pi\sum_iA_T(R_i)\Psi^*(\vec R_\alpha,\vec R_i)\Psi(\vec R_i,\vec R_\beta),
 \end{eqnarray}
    
 \noindent
 where ${\bf B_{\alpha \beta}}=\Psi(\vec R_\alpha,\vec R_\beta)$ and ${\bf A}$ is 
 the coupling matrix of trilobite states centered at different perturbers \cite{ours}. The off-diagonal
  elements are proportional to the overlap between trilobite wavefunctions localized on different 
  perturbers and account for the non-additive pair-wise nature of the Rydberg-neutral interaction 
  when multiple perturbers lie within the Rydberg orbit. Each of the $N$ nonzero eigenvalues 
  of this expression defines an $3N$-dimensional potential energy surface; the $k$th- eigenenergy 
  corresponds to the eigenstate
 
  \begin{align}
  \label{polytrilobite}
\Psi_k(\{\vec R_i\},\vec r) &= \frac{1}{\mathcal N}\sum_ia_{ik}\Psi(\vec R_i,\vec r);\\
\mathcal{N}^2&=\sum_{i,j}a_{ik}a_{jk}\Psi(\vec R_i,\vec R_j).\nonumber
 \end{align}
 For the single-perturber case the sole non-zero eigenenergy is given in closed form: \cite{ChibisovPRL}
 \begin{eqnarray}
 \label{trilobiteenergies}
E(R) &=2\pi A_T(R) \Psi(R,R)  \\
&=2\pi A_T(R)[u_{n0}(R)Y_{00}(\hat R)]^2\left([k(R)]^2 + [Q(R)]^2\right)\nonumber; 
 \end{eqnarray}
 
 \noindent
 where $Q(R)=u_{n0}'(R)/u_{n0}(R)$. The potential energy curves for a single perturber within 
 the Rydberg orbit are shown in Fig. \ref{Matt_figure}a, which shows the oscillatory structure of the potentials. Some of the oscillations exhibit local minima in $E(R)$ due to the energy dependent 
 nature of the phase-shift. 
 
We propose a two-photon 
 excitation scheme from the spin-polarized $1S$ ground state to either of the $nS$ or $nD$ Rydberg states, both of which are allowed by selection rules, using two counter-propagating, linearly-polarized beams with $\lambda \approx 181$ nm. With this excitation scheme the momentum kick of the photons to the atom will be negligible. The transition amplitudes for these two different routes, denoted as $P^{n}_{S}$ and $P^{n}_{D}$, respectively, have been calculated using a Sturmian basis set leading to accurate results (within $\sim10$\%) in comparison with other approaches  \cite{Quattropani-1982,Joshi-2006}. We have calculated $P^{20}_S = -0.11$, $P_D^{20} =0.48$; the ratio of these values changes by less than $10\%$ for $n = \infty$. We have therefore assumed these values for all $n$ \cite{Florescu}.  The trilobite state can be excited via its $nS$ or $nD$ character, depending on the nature of the Rydberg state. When a single perturber is present the probability of excitation, allowing for both pathways and averaged over the relative angle between the internuclear axis and the quantization axis, is:
 \begin{align}
 \label{onepertsprob}
\fl\mathcal{P}( R)&=(4\pi)^{-1}\int\dd{\Omega}\left|\bkt{nD}{\Psi_T}P^n_D + \bkt{nS}{\Psi_T}P^n_S\right|^2\nonumber\\
&=\frac{ (P_S^n)^2 +\left(\frac{u_{n2}(R)}{u_{n0}(R)}P_D^n\right)^2}{[R^2({[k(R)]^2+[Q(R)]^2})]}.
 \end{align}
  
     \begin{figure*}[t]
\begin{center}
\includegraphics[scale=0.7]{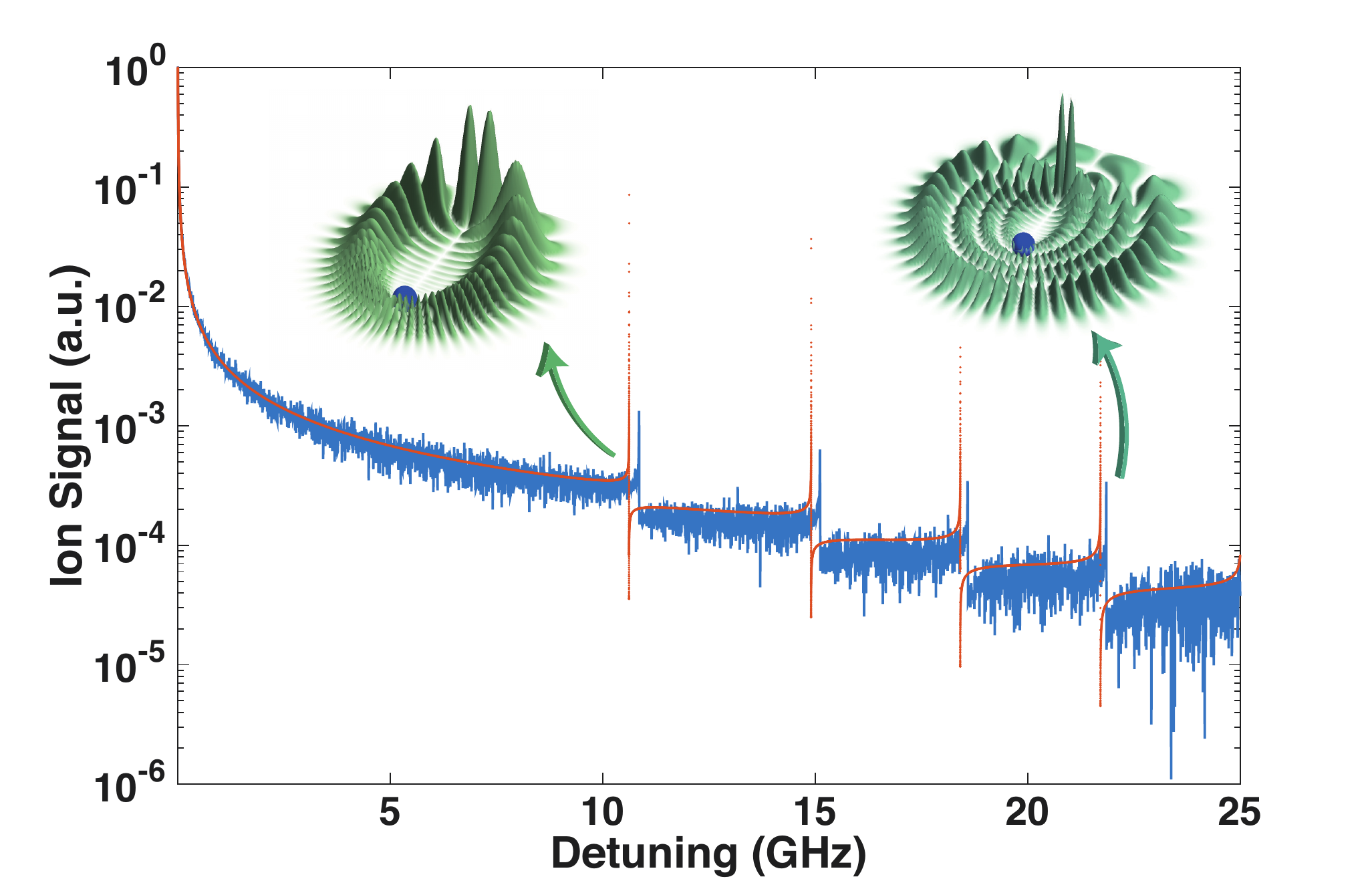}
\caption{Normalized spectra of a single hydrogen Rydberg atom with $n$ = 20 in a high density background gas of 
hydrogen at $\rho$ = 10$^{15}$ cm$^{-3}$. Only detunings higher than 25 MHz, where the quasistatic approximation is accurate, are shown. The quasistatic results for the line shape is depicted by the 
solid blue line, whereas the red line represents the line shape assuming a two-body quasistatic 
approach (see Eq. \ref{lineshapeanalyticsimple} in the text for details). Each of the peaks of the spectra correlates with 
the existence of local extrema in the Rydberg-perturber interaction potential.  $5\times 10^6$ Monte Carlo events have been employed in this simulation.}
\label{Spectrum}
\end{center}
\end{figure*}
\noindent The probability of excitation as a function of the distance for different Rydberg states are shown in Fig. \ref{Matt_figure}b. These show an oscillatory behavior related to the $nD$ and $nS$ hydrogenic wave 
functions, but tend towards a constant for large perturber-Rydberg distance.  For multiple perturbers, the probability to excite the $k$th eigenstate is given by 
 \begin{align}
 &\mathcal{P}^k(\{\vec R_i\})=\nonumber\\&\sum_ia_{ik}\left|P_S^n\bkt{nS}{\Psi(\vec R_i,\vec r)}+P_D^n\bkt{nD}{\Psi(\vec R_i,\vec r)}\right|^2  \label{Npertsprob}\nonumber\\&= \frac{\left[\sum_ia_{ik}\left(P_D^n\psi_{n20}(\vec R_i))+P_S^n\psi_{n00}(\vec r_i)\right)\right]^2}{\sum_{i,j}a_{ik}a_{jk}\Psi(\vec R_i,\vec R_j)}.
 \end{align}

\noindent The spectroscopy of a single Rydberg hydrogen atom in an ultracold background gas of 
 hydrogen is modeled assuming the quasistatic theory of line broadening \cite{Stellar,Kuhn-1934,Kuhn-1937,Allard-1982}. 
 In this theory, the absorber (the Rydberg atom) is assumed to be at rest and absorbs photons  
 at $\omega_{0} + \Delta \omega$, where $\omega_{0}$ is the frequency 
 between the two states of the Rydberg atom and $\Delta \omega$ accounts for the shift of the involved levels 
 by the perturbing potential of the neighboring atoms. An additional assumption is that only the perturbers 
 located in a given interacting region, which in 
 the present case is the volume of the Rydberg atom, $V_{Ryd}$, will contribute to the line broadening.

  The quasistatic picture for the line broadening breaks down when the collision time is close to or less than the inverse detuning, $\tau_{c} \sim 1/\Delta \omega$ 
 \cite{Allard-1982,Holstein-1950}. In this regime the Rydberg-perturber collisions give the major 
 contribution to the line broadening. The collision time can be estimated as $\tau_{c}=b/\langle v\rangle$, where 
 the impact parameter $b$ is assumed to be of the same order of magnitude as the Rydberg orbit 
 size, {\it i.e.}, $b \sim 2n^2$. Considering a thermal cloud of hydrogen at $T\sim$ 50 $\mu$K,  $\tau_{c} \sim$ 20 - 200 ns. Thus, for a thermal cloud of hydrogen the line shapes in the range
 $\Delta \omega \gtrsim $ 25 MHz can be approximated by by the quasistatic approach. 
     
The simulations of the line profile are performed following the method of Schlagm\"{u}ller et al. \cite{Schlag}. In a homogeneous gas of density $\rho$, the probability of finding a certain number 
of atoms in the volume of the Rydberg atom defines a Poisson distribution with a mean number $\langle N \rangle=\rho V_{Ryd}$, where $\langle N \rangle$ denotes the average number of perturbers in the
 Rydberg volume. This Poissonian distribution is employed to sample uniformly the $N$ atoms within the Rydberg volume. 
 For the $i$-th sample of the number of perturbers in the Rydberg orbit, the target Rydberg state experiences an energy 
 shift $E_j$; each of these energies are then weighted by the probability of the excitation of the 
 given configuration.
 Finally the line shape is given by the $\mathcal{P}$-weighted distribution of calculated energies in each of the samples, $S(E)$. In the present work the reported spectra are presented in terms of the ion signal since an ionization detection technique is assumed, as is commonly utilized in the field. The spectra are normalized to unity with respect to the highest signal at small detunings.

The spectrum of a single hydrogen Rydberg atom with $n = 20$ in a dense 
and ultracold gas is shown in Fig.\ref{Spectrum}, which prominently displays the existence of certain sharp spectral features
following a quasi-periodic pattern in terms of the detuning of the excitation field.  This pattern 
is intimately related with the underlying PEC; in particular, 
each of the peaks reflects the existence of a plateau in the potential energy landscape
 (see  Fig.\ref{Matt_figure}). In each of these plateaus the Rydberg electron is repeatedly elastically 
 scattered by, primarily, just one of the perturbing atoms, leading to a trilobite-like wave function for the Rydberg-perturber 
 system \cite{GreeneSadeghpourDickinson}. At each plateau, as the detuning increases, an angular node is exchanged for an additional radial node such that the total number of nodes remains a constant integer value at each plateau. This reflects the fact that, at each plateau, the electronic wave function is predominantly characterized by a single elliptical eigenstate \cite{Granger}. These wave functions are displayed in Fig. \ref{Spectrum}, where the Rydberg core
 is depicted by the blue ball whereas the perturber is placed underneath the electron density maximum. Therefore, the line shape of the Rydberg spectrum  directly 
maps the Rydberg-perturber PEC, leading to a very robust spectroscopic method for observation of this system.

 A simple model for the lineshape, assuming that only a single perturber lies in the Rydberg 
 orbit, conveys significant physical intuition about the system's lineshapes and spectral features. This has been shown in the case of 
 long-range forces by Kuhn \cite{Kuhn-1934,Kuhn-1937}. This assumption is clearly more 
 accurate for dilute gases or low Rydberg excitations, but is also a valuable limiting case 
 for the many-perturber scenario and qualitatively displays some of the same features. For a given density $\rho$ the probability to find a single perturber between distances $R$, $R + dR$ from the ion is given by the nearest neighbour distribution,
 \be
 \label{nnd}
 P(R) = \frac{3}{\Delta}\left(\frac{R}{\Delta}\right)^2e^{-(R/\Delta)^3},\,\,\,\,\Delta=(4\pi\rho/3)^{-1/3}.
 \ee
This distribution must be additionally modified to include the radial dependence of the radial probability of exciting the trilobite state, i.e. the probability distribution is that of Eq. \ref{nnd} multipled by Eq. \ref{onepertsprob}:
 \be
 P(R) = \frac{3e^{-(R/\Delta)^3\left[(P_S^n)^2+\left(P_D^n\frac{u_{n2}(R)}{u_{n0}(R)}\right)^2\right]}}{\Delta^3({[k(R)]^2+[Q(R)]^2})}\nonumber.
 \ee
The line shape is then given by converting this into a probability distribution with respect to the energy via the relationship $|P[R(E)]\dd{R}| = |P(E)\dd{E}|$. This introduces the derivative of the potential energy, which can be calculated analytically:
 \begin{align*}
 \frac{dE}{dR} &=\frac{dV(R)}{dR}=-A_T(R)\left(\frac{u_{n0}(R)}{R}\right)^2\\&\,\,\,\,\,\,\, +\frac{ A_T'(R)[u_{n0}(R)]^2}{2}\left([k(R)]^2 + [Q(R)]^2\right)+\frac{2\alpha_d}{R^5}.\nonumber
 \end{align*}
 Upon inverting the PEC to obtain $R$ as a function of the detuning, denoted $R_E = R(E)$, the line shape for the situation of a single perturbing atom within the Rydberg orbit is given by
 \begin{align}
\label{lineshapeanalytic}
 P(E) = \frac{3e^{-(R_E/\Delta)^3\left[(P_S^n)^2+\left(P_D^n\frac{u_{n2}(R_E)}{u_{n0}(R_E)}\right)^2\right]}}{\Delta^3({[k(R_E)]^2+[Q(R_E)]^2})}\left|\frac{dE}{dR_E}\right|^{-1}.
 \end{align}
This expression compactly separates into two factors. The first, a broad background, is given by the first factor of Eq. \ref{lineshapeanalyticsimple}. The second contains the peak structure and is unity for nearly all detunings except at the detunings of the series of sharp doublet peaks, and is given by the second factor of Eq. (\ref{lineshapeanalyticsimple}):
 \begin{align}
 \label{lineshapeanalyticsimple}
 P(E)  
 &= \left(\frac{3}{2}\frac{R_E^2e^{-(R_E/\Delta)^3}}{\Delta^3E}\right)\\&\cdot\left(\frac{\left[(P_S^n)^2+\left(P_D^n\frac{u_{n2}(R_E)}{u_{n0}(R_E)}\right)^2\right]}{1 - \frac{1}{A_T(R_E)[u_{n0}(R_E)]^2}\left(\frac{A_T'(R_E)}{A_T(R_E)}R_E^2 E + \frac{2\alpha_d}{R_E^3}\right)}\right)\nonumber.
 \end{align}
 
 \noindent
The smooth background given by the first factor can also be obtained by considering the approximate PEC derived by Borodin and Kazansky \cite{BandK}.  The asymptotic limits for small and large detunings of the functional form in Eq. (\ref{lineshapeanalyticsimple}) establishes the power-law scaling of the lineshapes, previously introduced by Kuhn for the case 
 of van der Waals forces \cite{Kuhn-1934,Kuhn-1937}. For small 
 detunings, corresponding to large $R$, the potential is very insensitive to $R$ and so the inverse function $R_E$ is nearly constant. The first factor in Eq. \ref{lineshapeanalyticsimple} is therefore proportional to $E^{-1}$, and the second is unity in this limit. A power law fit to the simulated lineshapes over an energy range from zero to the location of the first peaks gives $E^{-0.86\pm 0.01}$,  quite independently of density and the Rydberg state. This change in the power law, along with the small constant shift in the location of the doublet peak structure in Fig. \ref{Spectrum}, indicate the expected deviation beyond the single-perturber model.
 
 At large $E$, corresponding to fairly 
 low $R$, the radial wavefunction $[u_{n0}(R)]^2$ increases roughly proportional to $R^{1/2}$; the modulating 
 factor $k^2 + Q^2$ introduces a factor $R^{-1}$ making the leading order dependence of the potential curve 
 for small $R$ (but not so small as for the polarization potential to dominate yet) $E\sim R^{-1/2}$. Inverting
 then gives $R \sim E^{-2}$, so that $P(E)\sim E^{-5}$. This is a good estimate for
 the power law behavior in this regime, which numerical fits typically match well with $P(E) \sim E^{-4}-E^{-5}$, for detunings greater than $\sim 10$ GHz for $n=30$. This power law behavior is of course only satisfied when the second factor is unity; near the peak regions this term dominates. 
 
\begin{figure}[t]
\begin{center}
\includegraphics[scale=0.4]{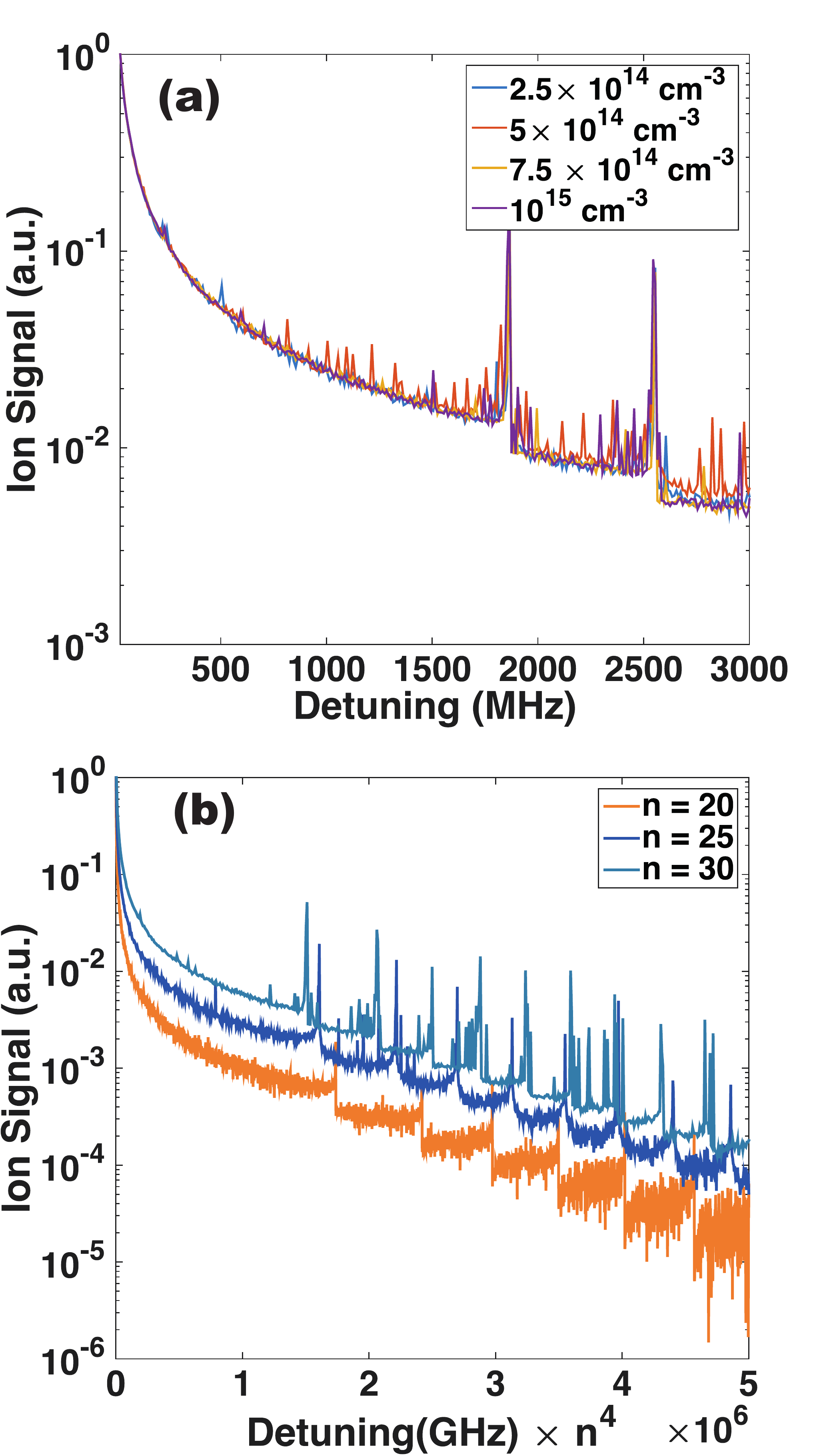}
\caption{Normalized spectra of a single hydrogen Rydberg atom in a thermal gas at ultracold temperatures. The Rydberg spectra for
 $n = 30$ and different densities is shown in panel (a). $10^6-5\times 10^5$ Monte Carlo events were simulated, with the number of events decreasing as the density increases. The spectrum for $\rho$ = 10$^{15}$ cm$^{-3}$ for different 
 Rydberg states is shown in panel (b). The horizontal axis has been scaled by $n^4$ as in Fig. 1 to emphasize the regular spacing of the peaks and the increased number of peaks per unit detuning as $n$ increases. $5\times 10^6$ events were simulated for $n = 20$ and $8\times 10^5$ for $n = 25,30$ }
\label{lineshape}
\end{center}
\end{figure}

An analysis of the components of equation (\ref{lineshapeanalytic}) explains the locations and shape of these sharp peaks in the spectra. For clarity throughout this discussion, we ignore the energy dependence of the scattering length and the polarization potential, so that $|dE/dR_E|^{-1} = |A_T(R)(u_{n0}(R)/R)^2|^{-1}$. This clearly demonstrates that the peaks in the spectrum stem from the plateaus of the potential energy curve. These are located at the nodes of the $s$-wave radial wave function. The first factor of eqn. (\ref{lineshapeanalytic}) also depends on $R_E$ in a complicated fashion, but near the nodes of $u_{n0}(R_E)$ it behaves as $[P_S^nu_{n0}(R_E)]^2 + [P_D^nu_{n2}(R_E)]^2$. If only the $S$ component was excited, the $u_{n0}(R_E)$ factors would cancel and the spectrum would no longer exhibit peaks. However, the $nD$ component is proportional to the ratio $u_{n2}(R_E)/u_{n0}(R_E)$ and thus vanishes at the node of the $d$-wave radial wave function very near the peak due to the node of the $s$-wave radial wave function, leading to the asymmetric profile seen in Figs. (\ref{Spectrum}) and (\ref{lineshape}) where a sharp peak is immediately followed by a step-like drop in the line shape. Inclusion of the energy dependence of the scattering length in fact allows the $nS$ character to exhibit peaks since it slightly offsets the location of the inflection points of the potential curve from the location of the $s$-wave radial nodes, but this effect is mostly overwhelmed by the dominant $nD$ character. 

The dependence of the lineshape on density is another fascinating feature of this system. The spectrum of a single Rydberg hydrogen atom with $n$ = 30 immersed in a perturbing gas 
as a function of its density is displayed in Fig.\ref{lineshape}a. The line shape is found to be 
independent of the density of the perturbing gas. These results seem to contradict the established 
line broadening theory which predicts a linear dependence of the line broadening with respect to 
the density of the perturbing gas \cite{Allard-1982}, since higher numbers of perturbers lead to larger shifts of the excited levels involved in the absorption process. However, the special 
nature of this system shows that the effect of different perturbers is clearly non-additive  \cite{ours}. 
In particular, for $N$ perturbers, $N$ eigenenergies of Eq. \ref{eq-1} split about the single-perturber eigenenergy, and so on average the contributions from higher-energy potential energy curves will be counteracted by contributions from lower-lying curves; as a result the line shape should be independent of the number of perturbing atoms and only depend on the Rydberg state. Indeed, this is numerically corroborated in Fig.\ref{lineshape}b, where 
the spectrum for a given density and different Rydberg states is shown. These results will no longer apply when the probability of finding at least one perturber within the Rydberg orbit is substantially smaller than one, or else so high that the number of perturbing atoms exceeds the number of states available to construct the trilobite state, $n^2$. These two limits set the range of applicable densities at $\langle N\rangle \sim 0.1 \le \frac{32\pi n^6}{3}\rho \le n^2$.  Within these limits, this system would be the first, to our knowledge, to exhibit a lineshape independent of the density.

The dependence on $n$ has also been explored: the
Rydberg spectrum of a single Rydberg excitation in
a dense background gas as a function of the Rydberg
state has been calculated within the quasistatic
approach of the line broadening and the results are
shown in Fig.3b. for three Rydberg states at the same density. These lineshapes possess the expected regular series of peaks, which can be correlated with those in Fig. 1. They show the same overall behavior as expected based on the potential energy curves, especially in that they exhibit the same regular scaling, $n^4$, as the potential energy curves. 

In this work a generalization of the quasistatic line broadening theory has been applied for the 
Rydberg excitation spectra of a single Rydberg hydrogen atom immersed in a high-density 
background hydrogen gas. The simulations not only account for the position and energy shift 
due to the perturbers, but also the probability of excitation of different atomic configurations 
by means of the $S$ and $D$ character of the target state.  As a result, Rydberg hydrogen atoms immersed in a dense background gas of hydrogen will show a quasi-periodic series of peaks 
in the line shape correlating with the fundamental nature of the underlying Rydberg-neutral
 interaction. In particular, the positions of these peaks relate closely to the locations of the plateaus in the potential curves, which depend sensitively on the calculated energy dependent electron-hydrogen scattering lengths. Thus, our 
 findings clearly indicate the possibility to use direct line shape data to explore the Rydberg-perturber
 energy landscape.

The calculated spectra clearly show a positive detuning in relation with the repulsive nature of the 
Rydberg-perturber interaction, as well as a density independent line shape, in stark contrast with 
the conventional line broadening theory, which predicts a linear density dependence of the 
line shift and line broadening. These findings have been explained by appealing to 
the unusual and intriguing properties of the polyatomic trilobite-like states, suggesting a 
consistent and convincing explanation of the spectra. 

Rydberg states are also present in astrophysical spectra, where even $n = 1000$ and higher have been observed \cite{Stebbings,biggest}. Although these Rydberg states form in very dilute interstellar clouds where the interatomic spacings are very large, trilobite states could form for very high $n$. The theory developed here could be applied to explain some aspects of the recombination lineshape due to the effect of a nearby neutral atom.

\ack
This work is supported in part by the National Science Foundation under Grant No. PHY-1306905. We appreciate the assistance and help of M. Schlagm\"{u}ller  in the many-body simulations for the line 
broadening, as well as fruitful discussions with F Robicheaux.
 
 \section*{References}

\end{document}